\documentclass[twocolumn,showpacs,prl,superscriptaddress]{revtex4}

\usepackage{mathrsfs}
\usepackage{amsmath}
\usepackage{amssymb}
\usepackage{graphicx}
\usepackage{amsfonts}
\usepackage{txfonts}

\begin{document}

\title{Dynamics revealed by correlations of time-distributed weak measurements of a single spin}

\author{R.-B. Liu}
\thanks{Corresponding author; rbliu@phy.cuhk.edu.hk}
\author{Shu-Hong Fung}
\author{Hok-Kin Fung}
\affiliation{Department of Physics, The Chinese University of Hong Kong, Shatin, N.T., Hong Kong, China}
\author{A. N. Korotkov}
\affiliation{Department of Electrical Engineering, University of California, Riverside, California 92521-0204}
\author{L. J. Sham}
\affiliation{Department of Physics, University of California San Diego, La Jolla, California 92093-0319}
\date{\today}

\begin{abstract}
We show that the correlations in stochastic outputs of time-distributed
weak measurements can be used to study the dynamics of an individual quantum object,
with a proof-of-principle setup based on small Faraday rotation caused by a
single spin in a quantum dot. In particular, the third order correlation can reveal
the ``true'' spin decoherence, which would otherwise be concealed by the
inhomogeneous broadening effect in the second order correlations. The viability of such approaches
lies in that (1) in weak measurement the state collapse which would disturb the system dynamics
occurs at a very low probability, and (2) a shot of measurement projecting the quantum object
to a known basis state serves as a starter or stopper of the evolution without pumping or
coherently controlling the system as otherwise required in conventional spin echo.
\end{abstract}

\pacs{76.70.Hb, 03.65.Ta, 42.50.Lc, 76.30.-v}

\maketitle

The standard von Neumann quantum measurement may be generalized in two aspects.
One is measurements distributed in time~\cite{Peres_Quantum_theory,Caves:1986PRD},
continuously or in a discrete sequence, as in the interesting Zeno~\cite{Peres_Quantum_theory}
and anti-Zeno effects~\cite{Kofman:2000}. Time-distributed measurements intrinsically interfere
with the evolution of the quantum object~\cite{Caves:1986PRD}. Another generalization is weak
measurement in which the probability of distinguishing the state of a quantum object by a single
shot of measurement is much smaller than one~\cite{Carmichael_open,wiseman93,Mensky_98,Korotkov_1999,Korotkov_2001}.
On the one hand, weak measurement has very low information yield rate; on the other hand,
it only rarely disturbs the dynamics of a quantum object by state collapse.
As a combination of the two generalizations, time-distributed weak measurements
have been used to steer the quantum state evolution~\cite{Katz:2006}.
In this paper, we show that the statistical analysis of time-distributed weak measurements
may be used to study the dynamics of a quantum object~\cite{Korotkov_2001}.
The outputs of time-distributed measurements bear the stochastic nature of quantum measurements,
so the standard noise analysis in quantum optics~\cite{Scully:quantumoptics} would be a
natural method to be applied. Notwithstanding that, we should emphasize that the stochastic output
of time-distributed weak measurement is not the noise in the system, but an intrinsic quantum
mechanical phenomenon. Revealing quantum dynamics by correlations of time-distributed
weak measurements is complementary to the fundamental dissipation-fluctuation theorem which
relates correlations of thermal noises to the linear response of a
system~\cite{Crooker:2004,Jiang_FET_spin,Oestreich:2005,Braun:2007PRB}.

To demonstrate the basic idea, we consider the monitoring of coherent Lamor precession and decoherence
of a single spin in a quantum dot, which is relevant to exploiting the spin coherence
in quantum technologies such as quantum computing~\cite{Koppens_T2,Marcus_T2,Gurudev:2005PRL,Greilich:2006}.
The difficulty of studying the spin decoherence lies in the fact that the ``true'' decoherence due to
quantum entanglement with environments is often concealed by the rapid ``phenomenological''
dephasing caused by inhomogeneous broadening in ensemble measurements (e.g., in a typical GaAs quantum dot,
the spin decoherence time is $\sim 10^{-6}$~sec, but the inhomogeneous broadening dephasing time is
$\sim 10^{-9}$~sec~\cite{Koppens:2008PRL,Marcus_T2,Greilich:2006,Gurudev:2005PRL,Mikkelsen:2007,Berezovsky:2008}).
Note that many single-spin experiments are still ensemble experiments with temporal repetition of measurements.
To resolve the spin decoherence excluding the inhomogeneous broadening effect, spin
echo~\cite{Koppens:2008PRL,Marcus_T2,Berezovsky:2008,ESR_silicon_T2,Abe_Si} and
mode-locking of spin frequency~\cite{Greilich:2006} have been invoked.
In this paper, we will show that the spin dynamics can be revealed in correlations of the stochastic outputs
of sequential weak probes. In particular, the third order correlation singles out the ``true'' spin
decoherence. Unlike conventional spin echo, the present method involves no
explicit pump or control of the spin but uses the state collapse
as the starter or stopper of the spin precession.

\begin{figure}[b]
\includegraphics[width=0.95\linewidth]{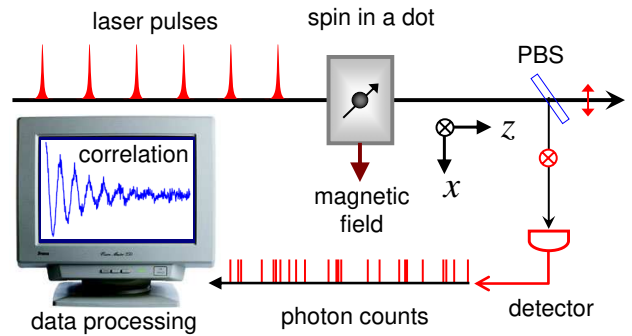}
\caption{A proof-of-principle setup for
weak measurement of a single spin in a quantum dot
by Faraday rotation. }
\label{setup}
\end{figure}

We design a proof-of-principle setup (see Fig.~\ref{setup}) based on
Faraday rotation, which has been used in experiments for spin
measurements~\cite{Greilich:2006,Berezovsky:2006,Mikkelsen:2007,Berezovsky:2008,Atature:2007NaturePhysics}.
The probe consists of a sequence of linearly polarized laser pulses
evenly spaced in delay time $\tau$. After interaction with a single spin
(in a quantum dot, e.g.), the light polarization is rotated by $\theta$ or
$-\theta$ for the spin state parallel or anti-parallel to
the light propagation direction ($z$-axis). The Faraday rotation angle $\theta$ by a single
electron spin is usually very small ($\sim 10^{-6}$~rad in a quantum
dot~\cite{Berezovsky:2006,Atature:2007NaturePhysics}), so the two polarization states
of the light corresponding to the two different spin states are almost identical.
Thus a detection of the light polarization is a weak measurement of the spin,
as long as the number of photons per pulse is not too large (see discussions following
Eq.~(\ref{distinguishability}) for details).
The light polarization is detected by filtering through a polarized beam splitter (PBS)
which is aligned to let the light with polarization rotated by $\theta$ fully pass through
and the light with orthogonal polarization be fully reflected.
The light with Faraday rotation angle $-\theta$ is reflected with probability $\sin^2(2\theta)$.
For a small $\theta$, the average number of reflected photons is much less than one, so in most cases,
a single-photon detector set at the reflection arm would be idle with no clicks and one cannot tell
which state the spin could be in. The clicks of the detector form a stochastic sequence.
The correlations in the sequence will be analyzed to study the spin dynamics,
such as the precession under a transverse magnetic field and the decoherence.
This proof-of-principle setup, being conceptually simple and adapted from
existing experiments, is of course not the only possible implementation.
For example, one can use continuous-wave probe instead of pulse sequences,
interferometer measurement of the polarization instead of the PBS filtering,
polarization-selective absorption instead of the Faraday rotation, and so on.

We shall derive from quantum optics description of the spin-light interaction
a weak measurement theory in the formalism of positive operator
value measure (POVM)~\cite{Peres_Quantum_theory,NielsenChuang}.
Consider a laser pulse in a coherent state $|\alpha,H\rangle \equiv e^{\alpha a_H^{\dag}-{\rm h.c.}}|0\rangle$
(where $a^{\dag}_{H/V}$ creates a photon with linear polarization $H$ or $V$)
and a spin in an arbitrary superposition $C_+|+\rangle+C_-|-\rangle$ in the basis
quantized along the $z$-axis, the initial spin-photon state is
\begin{equation}
|\psi\rangle=\left(C_+|+\rangle+C_-|-\rangle\right)\otimes |\alpha,H\rangle.
\end{equation}
After interaction, the state becomes an entangled one as
\begin{equation}
|\psi'\rangle=C_+|+\rangle\otimes|\alpha,{+\theta}\rangle+C_-|-\rangle\otimes |\alpha,{-\theta}\rangle,
\end{equation}
where $|\alpha,{\pm\theta}\rangle\equiv e^{\alpha a_{\pm\theta}^{\dag}-{\rm h.c.}}|0\rangle$
(with $a_{\pm\theta}\equiv a_{H}\cos\theta\pm a_{V}\sin\theta$)
is a photon coherent state with polarization rotated by $\pm\theta$.
How much the spin is measured is determined by the
distinguishability between the two polarization states
\begin{equation}
{\mathcal D}\equiv 1-\left|\langle\alpha,+\theta|\alpha,-\theta\rangle\right|^2
=1-\exp\left(-4|\alpha|^2\sin^2 \theta\right).
\label{distinguishability}
\end{equation}
When the average number of photons $\bar{N}=|\alpha|^2\gg 1$
and the Faraday rotation angle $\theta$ is not too small,
the two coherent states are almost orthogonal and ${\mathcal D}\rightarrow 1$,
thus a detection of the light polarization provides a
von Neumann projective measurement of the spin.
For a single spin in a quantum dot, the Faraday rotation angle $\theta$ is
usually very small. For example, in a GaAs fluctuation quantum dot~\cite{Berezovsky:2006},
$\left|\theta\right|\sim 10^{-5}~{\rm rad}$
for light tuned $1$~meV below the optical resonance with a focus spot
area $\sim 10~\mu$m$^{2}$. The number of photons in a 10~picosecond
pulse with power 10~mW is $\bar{N}\sim 0.5\times 10^{6}$. In this case,
${\mathcal D}\cong 4\bar{N}\theta^2\sim 2\times 10^{-4}\ll 1$, the spin states are almost indistinguishable
by the photon polarization states.
After interaction with the spin, the laser pulse is subject to the PBS
filtering which transforms the spin-photon state to be
\begin{align}
|\psi''\rangle=& C_+|+\rangle\otimes|\alpha\rangle_{t}\otimes
\left|0\right\rangle_{r}
+C_-|-\rangle\otimes
\left|\alpha\cos(2\theta)\right\rangle_{t}\otimes
\left|\alpha\sin(2\theta)\right\rangle_{r},
\end{align}
where $|\beta\rangle_{t/r}$ denotes a coherent state of the transmitted/reflected mode with amplitude $\beta$.
Separating the vacuum state $|0\rangle_r$ from the reflected mode and keeping terms
up to a relative error $O\left(\theta^2\right)$, we write the state as
\begin{align}
|\psi''\rangle= & \left(C_+|+\rangle+\sqrt{1-{\mathcal D}} C_-|-\rangle\right)\otimes|\alpha\rangle_{t}\otimes\left|0\right\rangle_{r}
\nonumber \\
&
+\sqrt{\mathcal D} C_-|-\rangle \otimes|\alpha\rangle_{t}\otimes \left|\alpha\sin(2\theta)\right\rangle'_{r},
\end{align}
where $\left|\alpha\sin(2\theta)\right\rangle'_{r}$ denotes the (normalized) state of the reflected mode but
with the vacuum component dropped. With a probability $P_1={\mathcal D}\left|C_-\right|^2\ll 1$,
an ideal detector at the reflection arm will detect a photon-click
and the spin state is known at $|-\rangle$, while in most cases (with probability $P_0=1-P_1$),
the detector will be idle and the spin state becomes $C_+|+\rangle+\sqrt{1-{\mathcal D}}C_-|-\rangle$
(up to a normalization factor), which is almost undisturbed by the measurement
[since the overlap between the state before the measurement and the state after the measurement
is $\left(|C_+|^2+\sqrt{1-{\mathcal D}}|C_-|^2\right)/\sqrt{1-|C_-|^2 {\mathcal D}}=1-O\left({\mathcal D}^2\right)$]. In the POVM
formalism~\cite{Peres_Quantum_theory,NielsenChuang}, the Kraus operators for the click and no-click cases
are respectively,
\begin{align}
\hat{M}_1  = \sqrt{\mathcal D}|-\rangle\langle -| , \ \  {\rm and} \ \
\hat{M}_0  = \sqrt{1-{\mathcal D}}|-\rangle\langle-|+|+\rangle\langle +|,
\end{align}
which determine the (non-normalized) post-measurement state $\hat{M}_{0/1}|\psi\rangle$ and
the probability $P_{0/1}=\left\langle\psi\left| \hat{M}^{\dag}_{0/1}\hat{M}_{0/1}\right|\psi\right\rangle$.


Between two subsequent shots of measurement, the spin precession under a transverse
magnetic field (along $x$-direction) is described by,
\begin{equation}
\hat{U}=\exp\left(-i\hat{\sigma}_x\omega\tau/2\right),
\end{equation}
where $\hat{\sigma}_x$ is the Pauli matrix along the $x$-direction, and
$\omega$ is the Larmor frequency.
Coupled to the environment and subject to dynamically fluctuating local fields,
the spin precession is always accompanied by decoherence. For simplicity, we
consider an exponential coherence decay characterized by a decoherence time $T_2$.
In the quantum trajectory picture~\cite{wiseman93,Scully:quantumoptics},
the decoherence can be understood as a result of continuous measurement
by the environment along the $x$-axis, for which the Kraus operators
for the quantum jumps with and without phase flip are respectively~\cite{NielsenChuang}
\begin{align}
\hat{E}_1  = \sqrt{\gamma/2} \hat{\sigma}_x, \ \ {\rm and} \ \ \
\hat{E}_0  = \sqrt{1-\gamma/2} \hat{I},
\end{align}
where $\gamma\equiv 1-\exp\left(-\tau/T_2\right)\cong \tau/T_2$ is the
coherence lost between two subsequent measurements.
For a spin state described by a density operator $\hat{\rho}$,
the decoherence within $\tau$ leads the state to $\hat{\mathscr E}[\hat{\rho}]
\equiv \hat{E}_0\hat{\rho}\hat{E}_0^{\dag}+\hat{E}_1\hat{\rho}\hat{E}_1^{\dag}$.

To study the spin dynamics under sequential measurement,
we generalize the POVM formalism for a sequence of $n$ measurement.
To incorporate the spin decoherence in the
density operator evolution, we define the superoperators for the weak measurement
and the free evolution as $\hat{\mathscr{M}}_{0/1} [\hat{\rho}]  = \hat{M}_{0/1} \hat{\rho}\hat{M}_{0/1}^{\dag}$,
$\hat{\mathscr{U}}[ \hat{\rho}]  = \hat{U} \hat{\rho} \hat{U}^{\dag}$,
in addition to the decoherence superoperator $\hat{\mathscr E}$ defined above.
For a sequence output $X\equiv [x_1x_2\cdots x_n]$ as a string of binary numbers,
the superoperator,
\begin{equation}
\hat{\mathscr M}_{X}=\hat{\mathscr M}_{x_n}\hat{\mathscr E}\hat{\mathscr U}\hat{\mathscr M}_{x_{n-1}}\cdots
\hat{\mathscr M}_{x_3}\hat{\mathscr E}\hat{\mathscr U}\hat{\mathscr M}_{x_2}\hat{\mathscr E}
\hat{\mathscr U}\hat{\mathscr M}_{x_1},
\end{equation}
transforms an initial density operator $\hat{\rho}$ to $\hat{\mathscr M}_{X}[\hat{\rho}]$
(not normalized) and determines the probability of the output $P_X={\rm Tr}\left(\hat{\mathscr M}_X[\hat{\rho}]\right)$.
With the POVM formalism, the spin state evolution under sequential measurement and hence the noise correlations
discussed below can be readily evaluated.

\begin{figure}[t]
\includegraphics[width=0.95\linewidth]{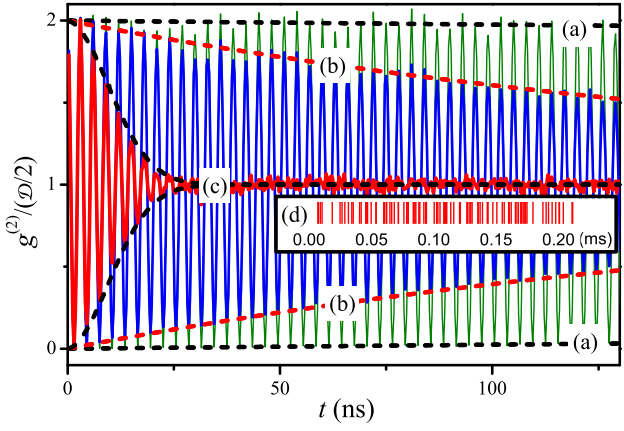}
\caption{The Monte Carlo simulation (solid oscillating curves)
and the analytical result (envelopes in dashed lines) of the 2nd order correlation
function, calculated with distinguishability ${\mathcal D}=3\times 10^{-4}$,
Lamor precession period $2\pi/\omega_0=3$~ns and the interval between two subsequent
measurements $\tau=0.3$~ns. In (a), no decoherence or inhomogeneous broadening
is present ($T_2^{-1}=\sigma=0$); In (b), $T_2=200$~ns but $\sigma=0$; In (c), $T_2=200$~ns and
$\sigma^{-1}=10$~ns. (d) shows the stochastic output (each line indicating a click event),
obtained in the Monte Carlo simulation of about $7\times 10^5$ shots of measurement during
a real time of about 0.2~ms, with the same condition as in (a).}
\label{g2_graph}
\end{figure}

To illustrate how a real experiment would perform,
we have carried out Monte Carlo simulations of the measurement with the following algorithm:
(1) We start from a randomly chosen state of the spin $|\psi\rangle$;
(2) The state after a free evolution is $\hat{U}|\psi\rangle$;
(3) Then the decoherence effect is taken into account by
    applying randomly the Kraus operator $\hat{E}_0$ or $\hat{E}_1$ to
    the state (with normalization) with probability $1-\gamma/2$
    or $\gamma/2$, respectively;
(4) The measurement is done by randomly applying the Kraus operator $\hat{M}_{0}$
    or $\hat{M}_1$ to the state (with normalization) corresponding to the output 0 or 1 (no-click or click),
    with probability $P_0$ or $P_1$ given by the POVM formalism.
Step (2)-(4) are repeated for many times.
The output is a random sequence of clicks, as shown in Fig.~\ref{g2_graph} (d).

To study the correlation of the stochastic output, we first consider the interval distribution $K(n)$,
defined as the probability of having two clicks separated by $n-1$ no-clicks~\cite{Scully:quantumoptics},
\begin{equation}
K(n)\equiv{\rm Tr}\left(\hat{\mathscr M}_{\left[10_{n-1}1\right]}[\hat{\rho}]\right)
 \left/{\rm Tr}\left(\hat{\mathscr M}_1[\hat{\rho}]\right)\right. ,
\end{equation}
where $0_{n-1}$ means a string of $n-1$ zeros. By a straightforward calculation,
\begin{align}
K(n)\approx & \frac{{\mathcal D}+{\mathcal D}^2}{2}e^{-\frac{n{\mathcal D}}{2}}
\left[1+e^{-\frac{n\tau}{T_2}} \cos
\left(n\omega \tau
 +\frac{{\mathcal D}}{2}\cot \frac{\omega\tau}{2}\right) \right],
\end{align}
up to $O\left(\gamma{\mathcal D}^2\right)$ and $O\left(n {\mathcal D}^3\right)$,
for $\gamma, {\mathcal D}\ll \omega\tau<\pi$.
A successful measurement at the beginning of an interval projects the spin to the basis state
$|-\rangle$ along the optical ($z$) axis. Then, the spin precesses under the external magnetic field
about the $x$-axis. The interval is terminated by a second successful measurement among the
periodic attempts after a time lapse of $n\tau$. The decay of the oscillation is due to
the spin decoherence. The overall decay $e^{-n{\mathcal D}/2}$ is due to decreasing of the probability
of unsuccessful measurement with increasing time. The measurement also induces a little
phaseshift to the oscillation. Obviously, the smaller the distinguishability
${\mathcal D}$, the less the spin dynamics is disturbed by the measurement.

In experiments, often the photon coincidence correlation instead of the interval distribution is measured.
The second order correlation $g^{(2)}(n\tau)$ is the probability of having two clicks separated
by $n-1$ measurements~\cite{Scully:quantumoptics}, regardless of the results in between,
\begin{align}
g^{(2)}(n\tau)  = &\sum_{x_1,x_2,\ldots,x_{n-1} \in\{0,1\}}{\rm Tr}\left(\hat{\mathscr M}_{1 x_1x_2\cdots x_{n-1} 1}[\hat{\rho}]\right)
/{\rm Tr}\left(\hat{\mathscr M}_1[\hat{\rho}]\right)
\nonumber \\
   = & K(n)+ \sum_{m=1}^{n-1}K(n-m)K(m)
 \nonumber \\
  & + \sum_{m=2}^{n-1}\sum_{l=1}^{m-1}K(n-m)K(m-l)K(l)+\cdots. \label{idf_series}
\end{align}
By Fourier transformation and summation in the frequency domain,
\begin{equation}
g^{(2)}(n\tau)=\frac{\mathcal D}{2}\left[
1+ e^{-n\left(\tau/T_2+{\mathcal D}/4\right)}
\cos\left(n\omega\tau\right)+O\left({\mathcal D}\right)\right]. \label{g2}
\end{equation}
The spin precession, the decoherence, and the measurement-induced decay are all
seen in the second order correlation function [see Fig.~\ref{g2_graph}].
Note that the overall decay of the interval distribution manifests itself as a
measurement-induced dephasing of the oscillating signal in the
correlation function.
The Monte Carlo simulation shows that $10^{10}$ shots of measurement would
yield a rather smooth profile of the spin dynamics, which requires a time span of
about 3~seconds for the parameters used in Fig.~\ref{g2_graph}.

In addition to the decoherence due to the dynamical fluctuation of the local field,
there is also phenomenological dephasing due to static or slow fluctuations, i.e.,
inhomogeneous broadening which exists even for a single spin since the sequential measurement
contains many shots which form an ensemble. The inhomogeneous broadening is
modeled by a Gaussian distribution of $\omega$ with mean value $\omega_0$ and width
 $\sigma$. With the inhomogeneous broadening included, the
ensemble-averaged correlation function becomes
\begin{equation}
\left\langle g^{(2)}(n\tau) \right\rangle
=\frac{\mathcal D}{2}\left[
1+ e^{-n\left(\tau/T_2+{\mathcal D}/4\right)-n^2\tau^2\sigma^2/2} \cos\left(n\omega_0\tau\right)
+O\left({\mathcal D}\right)\right]. \label{g2_IB}
\end{equation}
Since usually $\sigma\gg T_2^{-1}$, the decay of the 2nd order correlation is dominated by the
inhomogeneous broadening effect [see Fig.~\ref{g2_graph} (c)].

\begin{figure}[t]
\includegraphics[width=0.95\linewidth]{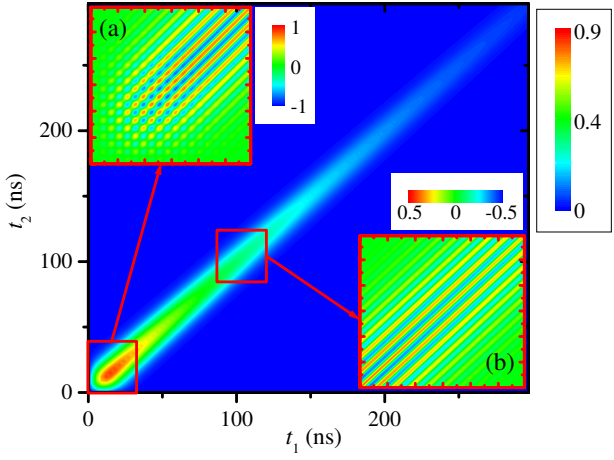}
\caption{Contour plot of the envelope of the 3rd
order correlation $G^{(3)}(t_1,t_2)$, with parameters the same as
for Fig.~\ref{g2_graph}~(c). The insets (a) and (b) show the
oscillation details of $G(t_1,t_2)$ in the range $0$~ns$\leq
t_{1,2} \leq 30$~ns and $90$~ns$\leq t_{1,2} \leq 120$~ns,
respectively.} \label{g3_contour}
\end{figure}

To separate the spin decoherence from the inhomogeneous broadening, we resort to
the 3rd order correlation $g^{(3)}(n_1\tau,n_2\tau)$, the probability of having
three clicks separated by $n_1-1$ and $n_2-1$ measurements. The idea can be understood
in a post-measurement selection picture: After the first click, the second click
has the peak probability appearing at an integer multiple of the spin precession
period, so the coincidence of the two earlier clicks serves as filtering of the spin
frequency and the third click would have a peak probability appearing at $n_2\tau=n_1\tau$,
similar to the spin echo. The 3rd order correlation in the absence of inhomogeneous
broadening is $g^{(3)}(t_1,t_2)\propto g^{(2)}(t_1)g^{(2)}(t_2)$. The ensemble-average
leads to
\begin{align}
\Big\langle & g^{(3)}(t_1,t_2) \Big\rangle  \propto
1+\sum_{j=1,2}e^{- \left(T_2^{-1}+\tau^{-1}{\mathcal D}/4\right)t_j
-\sigma^2 t_j^2/2} \cos\big({\omega}_0t_j\big)
\nonumber \\ &
 + \frac 1 2 e^{-\left(T_2^{-1}+\tau^{-1}{\mathcal D}/4\right)(t_1 + t_2)}e^{-\sigma^2
(t_1+t_2)^2/2} \cos\big(\omega_0(t_1+t_2)\big)
\nonumber \\ &
 + \frac 1 2 e^{-\left(T_2^{-1}+\tau^{-1}{\mathcal D}/4\right)(t_1 + t_2)} e^{-\sigma^2
(t_1-t_2)^2/2} \cos \big(\omega_0(t_1-t_2)\big).
\label{g3}
\end{align}
Fig.~\ref{g3_contour} plots
$G^{(3)}(t_1,t_2) \equiv \left\langle g^{(3)}(t_1,t_2)\right\rangle
 - \left\langle g^{(2)}(t_1) \right\rangle
   \left\langle g^{(2)}(t_2) \right\rangle$ to exclude the trivial background.
Along the direction $t_1=-t_2$, the 3rd order correlation oscillates and decays rapidly
(with timescale $\sigma^{-1}$). But the oscillation amplitude decays slowly
(with timescale $T_2$) along the direction $t_1=t_2$, as expected from the last term of Eq.~(\ref{g3}).

In conclusion, we have given a statistical treatment of sequential
weak measurements of a single spin. The characteristics of
the weak measurement consist in the negligible perturbation of the spin state except for
the projective state collapse when the measurement is successful in identifying the spin state.
We show that the third order correlation reveals the spin decoherence from the inhomogeneous broadening.
The theory presented here for sequential pulse measurement can be
straightforwardly generalized to continuous weak measurement by letting the pulse
separation $\tau\rightarrow 0$ while keeping the average power of the light
unchanged (i.e., ${\mathcal D}/\tau={\rm constant}$). In the proof-of-principle setup based on Faraday rotation,
all optical elements have been assumed ideal for conceptual simplicity.
An investigation of the defects, e.g., in the PBS and in
the photon detector, shows that they do not change the essential results presented here but only
reduce the visibility of the features. Details will be published elsewhere.

This work was supported by the Hong Kong RGC Project CUHK402207,
NSA/IARPA/ARO grant W911NF-08-1-0336, and ARO/LPS.


\end{document}